# Degree based Topological indices of Hanoi Graph

**Raghisa Khalid[1,*], Nazeran Idrees[1]**

[1]Faculty of Science and Technology, Department of Mathematics, Government College University, 38000 Faisalabad, Pakistan; nazeranjawwad@gmail.com
[*]Correspondence: kraghisa@yahoo.com; Tel,: +92333-6503299

**Abstract**

There are various topological indices for example distance based topological indices and degree based topological indices etc. In QSAR/QSPR study, physiochemical properties and topological indices for example $ABC$ index, $ABC_4$ index, Randic connectivity index, sum connectivity index, and so forth are used to characterize the chemical compound. In this paper we computed the edge version of $ABC$ index, $ABC_4$ index, Randic connectivity index, sum connectivity index, $GA$ index and $GA_5$ index of Double-wheel graph $DW_n$ and Hanoi graph $H_n$. The results are analyzed and the general formulas are derived for the above mentioned families of graphs.

## 1. Introduction and preliminary results

A single number, in graph theoretical terms, representing a chemical structure, is named as topological descriptor. A topological descriptor when correlates with a molecular property, it can be demonstrated as molecular index or topological index. Good correlation with the structure was found between the molecular properties for: thermodynamic properties (for example boiling points, heat of combustion, enthalpy of formation, etc.) and several boiling properties. Consequently, a topological index renovates a chemical structure into a particular number, beneficial in QSPR/QSAR studies.

In this paper all molecular graphs are considered to be connected, finite, loopless and deprived of parallel edges. . Let $F$ be a graph with n vertices and m edges. The degree of a vertex is the number of vertices adjacent to $q$ and is signified as $d(q)$. By these terminologies, certain topological indices are well-defined in the following way.

The oldest degree based topological index is Randic index signified as $\chi(F)$ and presented by Randic [11]. He suggested this index for calculating the degree of branching of the carbon-atom skeleton of saturated hydrocarbons.

**Definition 1.** For a molecular graph $F$, the Randic index is defined as

$$\chi(F) = \sum_{qr \in E(F)} \frac{1}{\sqrt{d_q d_r}}.$$

There is a relationship among Randic index and certain physiochemical properties of alkanes: surface area, boiling points, energy level, etc.

A variation of Randic connectivity index is the sum-connectivity index. It was presented by Zhou and Trinajstic [15]. They determined upper and lower bounds of this index for trees in terms of other graph invariants.

**Definition 2.** For a molecular graph $F$, the sum-connectivity index is defined as

$$S(F) = \sum_{qr \in E(F)} \frac{1}{\sqrt{d_q + d_r}}.$$

Estrada *et al.* in [3] proposed a degree based topological index of graphs, which is said to be atom-bond connectivity index. It can be used as tool to model the thermodynamic properties of organic compounds.

**Definition 3.** Let $F$ be molecular graph, then $ABC$ index is defined as

$$ABC(F) = \sum_{qr \in E(F)} \sqrt{\frac{d_q + d_r - 2}{d_q d_r}}.$$

Geometric-artithmetic index is associated with a variation of physiochemical properties. It can be used as possible tool for QSPR/QSAR research. Vukicevic and Furtula in [14] introduced the geometric-arithmetic $(GA)$ index.

**Definition 4.** Let $F$ be molecular graph, then geometric-arithmetic index is defined as

$$GA(F) = \sum_{qr \in E(F)} \frac{2\sqrt{d_q d_r}}{d_q + d_r}.$$

Ghorbani and Hosseinzadeh in [7] presented the fourth atom-bond connectivity index.

**Definition 5.** Let $F$ be molecular graph, then $ABC_4$ index is defined as

$$ABC_4(F) = \sum_{qr \in E(F)} \sqrt{\frac{S_q + S_r - 2}{S_q S_r}}.$$

Where $S_q$ is the summation of degrees of all the neighbors of vertex $r$ in $F$.

Recently Graovac *et al.* in [9] proposed fifth $GA$ index. As an essential topological index, the fifth geometric index is used to check the chemical properties of chemical compounds, nanomaterial and drugs.

**Definition 6.** Let $F$ be molecular graph, then $GA_5$ index is defined as

$$GA_5(F) = \sum_{qr \in E(F)} \frac{2\sqrt{S_q S_r}}{(S_q + S_r)}.$$

Das and Trinajstic in [2] associated the $ABC$ and $GA$ indices for molecular graphs and chemical trees also compared there two indices for general graphs. Later on Gan *et al.* in [8] introduced some sharp lower and upper bounds on $ABC$. Chen and Li in [1] gave sharp lower bound for sum-connectivity index having $n$-vertex unicyclic graphs by $k$ pendant vertices. Farhani in [4] investigated numerous topological indices in polyhex nanotubes: Randic connectivity index, sum connectivity index, atom-bond connectivity index, geometric-arithmetic index, first and second Zagreb indices and Zagreb polynomials. After that Farahani in [5] proposed $ABC_4$ index for Nanotori and V-Phenylenic Nanotube. After that Farhani in [6] gave explicit formulas for $GA_5$ index of a family of Hexagonal Nanotubes namely: Armchair Polyhex Nanotubes. Later on Sridhara *et al.* in [12] computed Randic index, $ABC$ index, $ABC_4$ index, sum connectivity index, geometric-arithmetic index and $GA_5$ index of Graphene.

Inspired by recent work on Graphene (Sridhara *et al.*, 2015) of computing topological indices, Shigehlli and Kanabur in [13] proposed new topological indices, namely, Arithmetic-Geometric index ($AG_1$ index), $SK$ index, $SK_1$ index and $SK_2$ index of a molecular graph G and found the explicit formulae of these indices for Graphene. After that Kanna *et al.* in [10] investigated Randic, ABC, sum connectivity, $ABC_4$, $GA$ connectivity and $GA_5$ indices of Dutch windmill graph.

## 2. Main results for Double-wheel graph

A double-wheel graph $DW_n$ of size n can be composed of $2C_n + K_1, n \geq 3$, that is it contains two cycles of size $n$, where all the points of the two cycles are associated to a common center.

The degree based topological indices like Randic, sum, atom-bond, geometric-arithmetic, fourth atom-bond, $GA_5$ index for Double-wheel graph are calculated in this section.

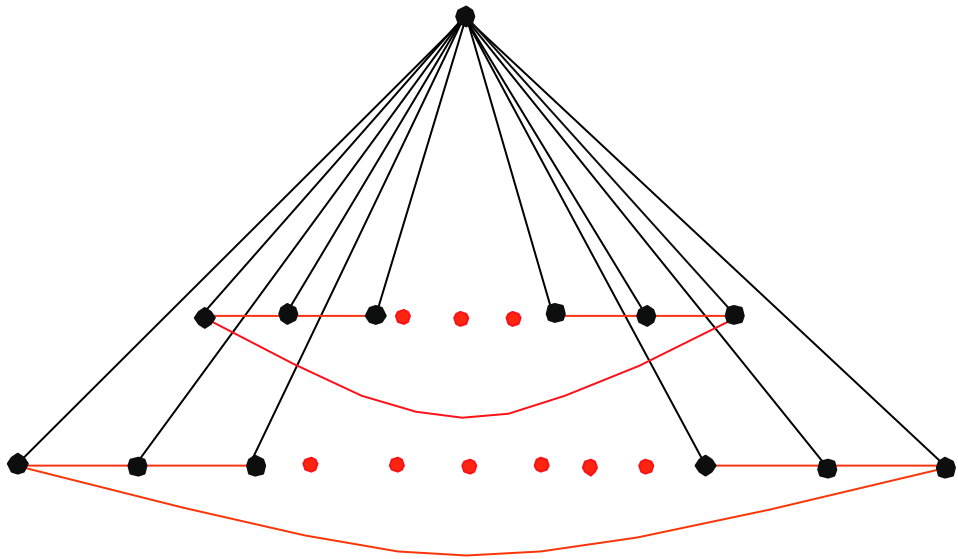

Figure 1: A representation of Double-wheel graph $DW_n$.

**Theorem 1**

The Randic connectivity index of Double-wheel graph is $\chi(DW_n) = \dfrac{2n}{3} + \dfrac{2n}{\sqrt{6n}}$.

Proof. Consider the Double-wheel graph $DW_n$. The edges of $DW_n$ can be partition into edges of form $E_{(d_u,d_v)}$, where $uv$ is an edge. We develop the edges of the form $E_{(3,3)}$ and $E_{(3,2n)}$. In figure 1 $E_{(3,3)}$ is colored in red. These forms for the sum of edges are given in the table 1.

We know that $\chi(F) = \sum\limits_{qr \in E(F)} \dfrac{1}{\sqrt{d_q d_r}}$.

$$\chi(DW_n) = |E_{(3,3)}| \sum\limits_{qr \in E_{(3,3)}(F)} \dfrac{1}{\sqrt{d_q d_r}} + |E_{(3,2n)}| \sum\limits_{qr \in E_{(3,2n)}(F)} \dfrac{1}{\sqrt{d_q d_r}}$$

From table 1 and figure 1,

$$= \dfrac{2n}{\sqrt{9}} + \dfrac{2n}{\sqrt{6n}}$$

$$\chi(DW_n) = \dfrac{2n}{3} + \dfrac{2n}{\sqrt{6n}}.$$

Here the proof of the theorem 1 is completed.

**Table 1:** Edge partition created by sum of adjacent vertices of every line.

| Edge of the form $E_{d_q,d_r}$ | Sum of edges |
|---|---|
| $E_{(3,3)}$ | $2n$ |
| $E_{(3,2n)}$ | $2n$ |

**Theorem 2**

The sum connectivity index of Double-wheel graph is $S(DW_n) = \dfrac{2n}{\sqrt{6}} + \dfrac{2n}{\sqrt{3+2n}}$.

Proof. We know that $S(F) = \sum\limits_{qr \in E(F)} \dfrac{1}{\sqrt{d_q + d_r}}$

$$S(DW_n) = |E_{(3,3)}| \sum\limits_{qr \in E_{(3,3)}(F)} \dfrac{1}{\sqrt{d_q + d_r}} + |E_{(3,2n)}| \sum\limits_{qr \in E_{(3,2n)}(F)} \dfrac{1}{\sqrt{d_q + d_r}}$$

From table 1 and figure 1,

$$= \frac{2n}{\sqrt{6}} + \frac{2n}{\sqrt{3+2n}}$$

$$S(DW_n) = \frac{2n}{\sqrt{6}} + \frac{2n}{\sqrt{3+2n}}.$$

Here the proof of the theorem 2 is completed.

**Theorem 3**

The $ABC$ index of Double-wheel graph is $ABC(DW_n) = \frac{4n}{3} + 2n\sqrt{\frac{1+2n}{6n}}$.

Proof. We know that $ABC(F) = \sum_{qr \in E(F)} \sqrt{\frac{d_q + d_r - 2}{d_q d_r}}$.

$$ABC(DW_n) = |E_{(3,3)}| \sum_{qr \in E(3,3)(F)} \sqrt{\frac{d_q + d_r - 2}{d_q d_r}} + |E_{(3,2n)}| \sum_{qr \in E(3,2n)(F)} \sqrt{\frac{d_q + d_r - 2}{d_q d_r}}$$

From table 1 and figure 1,

$$= 2n\sqrt{\frac{3+3-2}{9}} + 2n\sqrt{\frac{3+2n-2}{6n}}$$

$$= 2n\sqrt{\frac{4}{9}} + 2n\sqrt{\frac{1+2n}{6n}}$$

$$ABC(DW_n) = \frac{4n}{3} + 2n\sqrt{\frac{1+2n}{6n}}.$$

Here the proof of the theorem 3 is completed.

**Theorem 4**

The $GA$ index of Double-wheel graph is $GA(DW_n) = 2n + \frac{4n\sqrt{6n}}{3+2n}$.

Proof. We know that $GA(F) = \sum_{qr \in E(F)} \frac{2\sqrt{d_q d_r}}{d_q + d_r}$.

$$GA(DW_n) = |E_{(3,3)}| \sum_{qr \in E_{(3,3)}(F)} \frac{2\sqrt{d_q d_r}}{d_q + d_r} + |E_{(3,2n)}| \sum_{qr \in E_{(3,2n)}(F)} \frac{2\sqrt{d_q d_r}}{d_q + d_r}$$

From table 1 and figure 1,

$$GA(DW_n) = 2n + \frac{4n\sqrt{6n}}{3+2n}.$$

Here the proof of the theorem 4 is completed.

**Theorem 5**

The $ABC_4$ index of Double-wheel graph is $ABC_4(DW_n) = \dfrac{4n}{3} + 2n\sqrt{\dfrac{1+2n}{6n}}$.

Proof. Consider the Double-wheel graph $DW_n$. The edges of $DW_n$ can be partition into edges of form $E_{d_q,d_r}$, where $qr$ is an edge. We develop the edges of the form $E_{(2n+6,2n+6)}$ and $E_{(2n+6,6n)}$ that are shown in table 9. For convenience these edge kinds are colored by bright green and blue respectively, as shown in figure 2.

**Table 2:** Edge partition created by sum of degrees of neighbors of the head-to-head vertices of every edge.

| Edge of the form $E_{d_q,d_r}$ | Sum of edges |
|---|---|
| $E_{(2n+6,2n+6)}$ | $2n$ |
| $E_{(2n+6,6n)}$ | $2n$ |

We know that $ABC_4(F) = \displaystyle\sum_{qr \in E(F)} \sqrt{\dfrac{S_q + S_r - 2}{S_q S_r}}$.

$$ABC_4(DW_n) = \left|E_{(2n+6,2n+6)}\right| \sum_{qr \in E_{(2n+6,2n+6)}(F)} \sqrt{\dfrac{S_q + S_r - 2}{S_q S_r}} + \left|E_{(2n+6,6n)}\right| \sum_{qr \in E_{(2n+6,6n)}(F)} \sqrt{\dfrac{S_q + S_r - 2}{S_q S_r}}$$

From table 2 and figure 2,

$$= 2n\sqrt{\dfrac{2n+6+2n+6-2}{(2n+6)(2n+6)}} + 2n\sqrt{\dfrac{2n+6+6n-2}{(2n+6)(6n)}}$$

$$= \dfrac{2n}{2n+6}\sqrt{4n+10} + \dfrac{2n.2\sqrt{2n+1}}{2.\sqrt{3n^2+9n}}$$

$$ABC_4(DW_n) = \dfrac{2n}{2n+6}\sqrt{4n+10} + 2n\sqrt{\dfrac{2n+1}{3n^2+9n}}.$$

Here the proof of the theorem 5 is completed.

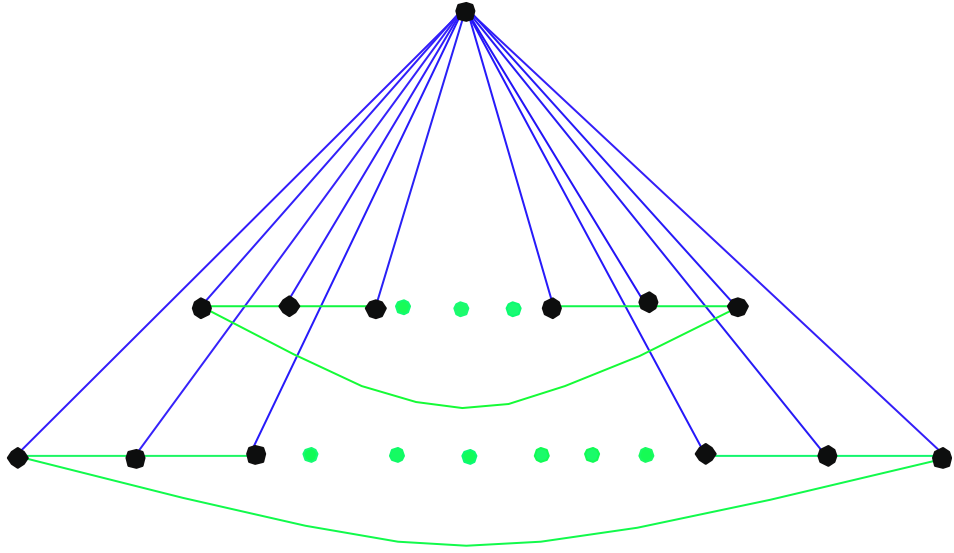

Figure 2

**Theorem 6**

The $GA_5$ index of Double-wheel graph is $GA_5(DW_n) = 2n + \dfrac{4n\sqrt{3n^2+9n}}{4n+3}$.

Proof. We know that $GA_5(F) = \displaystyle\sum_{qr \in E(F)} \dfrac{2\sqrt{S_q S_r}}{S_q + S_r}$.

$$GA_5(DW_n) = |E_{(2n+6, 2n+6)}| \sum_{qr \in E_{(2n+6, 2n+6)}(F)} \dfrac{2\sqrt{S_q S_r}}{S_q + S_r} + |E_{(2n+6, 6n)}| \sum_{qr \in E_{(2n+6, 6n)}(F)} \dfrac{2\sqrt{S_q S_r}}{S_q + S_r}$$

From table 2 and figure 2,

$$= 2n \dfrac{2\sqrt{(2n+6)^2}}{2n+6+2n+6} + 2n \dfrac{2\sqrt{12n^2+36n}}{2n+6+6n}$$

$$= \dfrac{4n(2n+6)}{2(2n+6)} + \dfrac{8n\sqrt{3n^2+9n}}{2(4n+3)}.$$

$$GA_5(DW_n) = 2n + \dfrac{4n\sqrt{3n^2+9n}}{4n+3}$$

Here the proof of the theorem 6 is completed.

### 3. Main results for Hanoi graph

The Hanoi graph $H_n$ can be created by taking the vertices to be the odd binomial coefficients of pascal's triangle calculated on the integers from 0 to $2^n - 1$ and drawing a line when coefficients are together diagonally or horizontally. The graph $H_n$ has $3^n$ vertices and

$\frac{3(3^n-1)}{2}$ edges. Every Hanoi graph has a unique Hamiltonian cycle.

The degree based topological indices like Randic, sum, atom-bond, geometric-arithmetic, fourth atom-bond, $GA_5$ index for Hanoi graph $H_n$ are computed in this section.

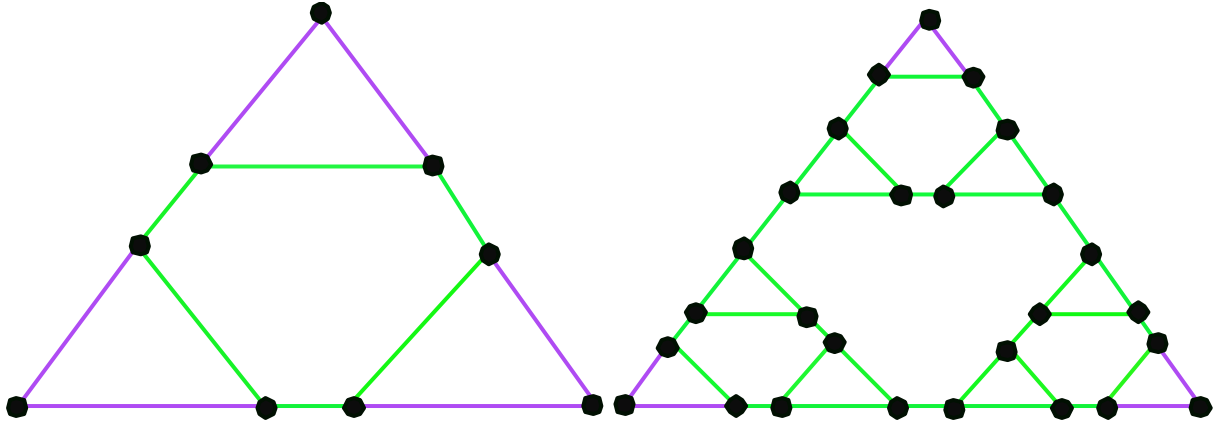

Fig 3: A representation of Hanoi graph.

**Theorem 7**

The Randic connectivity index of Hanoi graph is $\chi(H_n) = \sqrt{6} + \frac{3^{n+1}}{6} - \frac{5}{2}$.

Proof. Consider the Hanoi graph $H_n$. We partition the edges of $H_n$ into edges of form $E_{d_q,d_r}$, where $qr$ is an edge. We develop the edges of the form $E_{(2,3)}$ and $E_{(3,3)}$. In figure 4, $E_{(2,3)}$ and $E_{(3,3)}$ are colored in lavender and bright green, respectively. The sum of edges of these forms is given in the table 3.

We know that $\chi(F) = \sum_{qr \in E(F)} \frac{1}{\sqrt{d_q d_r}}$.

$$\chi(H_n) = |E_{(2,3)}| \sum_{qr \in E_{(2,3)}(F)} \frac{1}{\sqrt{d_q d_r}} + |E_{(3,3)}| \sum_{qr \in E_{(3,3)}(F)} \frac{1}{\sqrt{d_q d_r}}$$

From table 3 and figure 3,

$$= \frac{6}{\sqrt{6}} + \frac{3^{n+1} - 15}{2} \times \frac{1}{\sqrt{9}}$$

$$= \sqrt{6} + \frac{3^{n+1} - 15}{6}$$

$$\chi(H_n) = \sqrt{6} + \frac{3^{n+1}}{6} - \frac{5}{2}.$$

Here the proof of the theorem 7 is completed.

**In general case for $H_n$, when $n \geq 3$:**

**Table 3:** Edge partition created by sum of adjacent vertices of every line.

| Edge of the form $E_{d_q, d_r}$ | Sum of edges |
|---|---|
| $E_{(2,3)}$ | 6 |
| $E_{(3,3)}$ | $\dfrac{3^{n+1} - 15}{2}$ |

**Theorem 8**

The Sum-connectivity index of Hanoi graph is $S(H_n) = \dfrac{6}{\sqrt{5}} + \dfrac{3^{n+1} - 15}{2\sqrt{6}}$.

Proof. We know that $S(F) = \sum\limits_{qr \in E(F)} \dfrac{1}{\sqrt{d_q + d_r}}$.

$$S(H_n) = |E_{(2,3)}| \sum\limits_{qr \in E_{(2,3)}(F)} \frac{1}{\sqrt{d_q + d_r}} + |E_{(3,3)}| \sum\limits_{qr \in E_{(3,3)}(F)} \frac{1}{\sqrt{d_q + d_r}}$$

From table 3 and figure 3,

$$= \frac{6}{\sqrt{2+3}} + \frac{3^{n+1} - 15}{2} \times \frac{1}{\sqrt{3+3}}$$

$$S(H_n) = \frac{6}{\sqrt{5}} + \frac{3^{n+1} - 15}{2\sqrt{6}}.$$

Here the proof of the theorem 8 is completed.

**Theorem 9**

The $ABC$ index of Hanoi graph is $ABC(H_n) = 3\sqrt{2} + 3^n - 5$.

Proof. We know that $ABC(F) = \sum_{qr \in E(F)} \sqrt{\dfrac{d_q + d_r - 2}{d_q d_r}}$.

$$ABC(H_n) = |E_{(2,3)}| \sum_{qr \in E_{(2,3)}(F)} \sqrt{\dfrac{d_q + d_r - 2}{d_q d_r}} + |E_{(3,3)}| \sum_{qr \in E_{(3,3)}(F)} \sqrt{\dfrac{d_q + d_r - 2}{d_q d_r}}$$

From table 3 and figure 3,

$$= 6\sqrt{\dfrac{2+3-2}{6}} + \dfrac{3^{n+1} - 15}{2} \sqrt{\dfrac{3+3-2}{9}}$$

$$= \dfrac{6}{\sqrt{2}} + \dfrac{3^{n+1} - 15}{2} \sqrt{\dfrac{4}{9}}$$

$$= \dfrac{6}{\sqrt{2}} + \dfrac{3^{n+1} - 15}{3}$$

$$= \dfrac{6}{\sqrt{2}} + \dfrac{3^{n+1}}{3} - \dfrac{15}{3}$$

$$ABC(H_n) = 3\sqrt{2} + 3^n - 5.$$

Here the proof of the theorem 9 is completed.

**Theorem 10**

The $GA$ index of Hanoi graph is $GA(H_n) = \dfrac{12\sqrt{6}}{5} + \dfrac{3^{n+1} - 15}{2}$.

Proof. We know that $GA(F) = \sum_{qr \in E(F)} \dfrac{2\sqrt{d_q d_r}}{d_q + d_r}$.

$$GA(H_n) = |E_{(2,3)}| \sum_{qr \in E_{(2,3)}(F)} \dfrac{2\sqrt{d_q d_r}}{d_q + d_r} + |E_{(3,3)}| \sum_{qr \in E_{(3,3)}(F)} \dfrac{2\sqrt{d_q d_r}}{d_q + d_r}$$

From table 3 and figure 3,

$$= 6\frac{2\sqrt{6}}{5} + \frac{3^{n+1}-15}{2} \times \frac{2\sqrt{9}}{6}$$

$$GA(H_n) = \frac{12\sqrt{6}}{5} + \frac{3^{n+1}-15}{2}.$$

Here the proof of the theorem 10 is completed.

**Theorem 11**

The $ABC_4$ index of Hanoi graph is $ABC_4(H_n) = 3\sqrt{\frac{7}{32}} + 6\sqrt{\frac{5}{24}} + \frac{2(3^{n+1})}{9} - \frac{13}{3}$.

Proof. Consider the Hanoi graph $H_n$. We partition the edges of $H_n$ into edges of form $E_{d_q,d_r}$, where $qr$ is an edge. We develop the edges of the form $E_{(6,8)}$, $E_{(8,8)}$, $E_{(9,8)}$ and $E_{(9,9)}$ that are shown in table 4. For convenience these edge kinds are colored by green, blue, pink and red, respectively, as shown in figure 4.

**Table 4:** Edge partition created by sum of degrees of neighbors of the head-to-head vertices of every edge.

| Edge of the form $E_{d_q,d_r}$ | Sum of edges |
| --- | --- |
| $E_{(6,8)}$ | 6 |
| $E_{(8,8)}$ | 3 |
| $E_{(9,8)}$ | 6 |

We know that $ABC_4(F) = \sum_{qr \in E(F)} \sqrt{\frac{S_q + S_r - 2}{S_q S_r}}$.

$$ABC_4(H_n) = |E_{(6,8)}| \sum_{qr \in E_{(6,8)}(F)} \sqrt{\frac{S_q + S_r - 2}{S_q S_r}} + |E_{(8,8)}| \sum_{qr \in E_{(8,8)}(F)} \sqrt{\frac{S_q + S_r - 2}{S_q S_r}}$$

$$+ |E_{(9,8)}| \sum_{qr \in E_{(9,8)}(F)} \sqrt{\frac{S_q + S_r - 2}{S_q S_r}} + |E_{(9,9)}| \sum_{qr \in E_{(9,9)}(F)} \sqrt{\frac{S_q + S_r - 2}{S_q S_r}}.$$

From table 4 and figure 4,

$$= 6\sqrt{\frac{6+8-2}{48}} + 3\sqrt{\frac{8+8-2}{64}} + 6\sqrt{\frac{9+8-2}{72}} + \frac{3^{n+1}-33}{2}\sqrt{\frac{9+9-2}{81}}$$

$$= 6\sqrt{\frac{12}{48}} + 3\sqrt{\frac{14}{64}} + 6\sqrt{\frac{15}{72}} + \frac{3^{n+1}-33}{2}\sqrt{\frac{16}{81}}$$

$$= 3 + 3\sqrt{\frac{7}{32}} + 6\sqrt{\frac{5}{24}} + \frac{2(3^{n+1}-33)}{9}$$

$$= 3\sqrt{\frac{7}{32}} + 6\sqrt{\frac{5}{24}} + \frac{2(3^{n+1})}{9} - \frac{22}{3} + 3$$

$$ABC_4(H_n) = 3\sqrt{\frac{7}{32}} + 6\sqrt{\frac{5}{24}} + \frac{2(3^{n+1})}{9} - \frac{13}{3}.$$

Here the proof of the theorem 11 is completed.

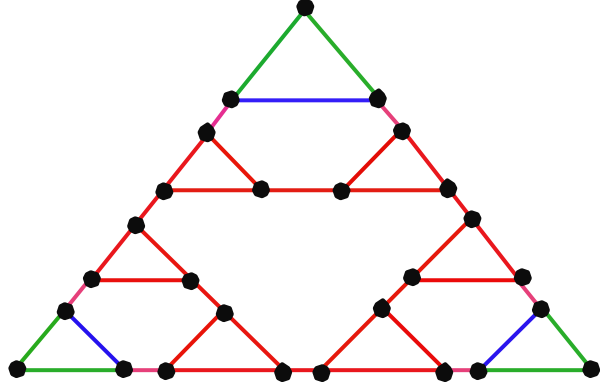

Figure 4

**Theorem 12**

The fifth geometric-arithmrtic connectivity index of Hanoi graph is

$$GA_5(H_n) = \frac{12^{3/2}}{7} + 3 + \frac{72\sqrt{2}}{17} + \frac{3^{n+1} - 33}{2}.$$

Proof. We know that $GA_5(F) = \sum_{qr \in E(F)} \frac{2\sqrt{S_q S_r}}{S_q + S_r}$.

$$GA_5(H_n) = |E_{(6,8)}| \sum_{qr \in E_{(6,8)}(F)} \frac{2\sqrt{S_q S_r}}{S_q + S_r} + |E_{(8,8)}| \sum_{qr \in E_{(8,8)}(F)} \frac{2\sqrt{S_q S_r}}{S_q + S_r}$$

$$+ |E_{(9,8)}| \sum_{qr \in E_{(9,8)}(F)} \frac{2\sqrt{S_q S_r}}{S_q + S_r} + |E_{(9,9)}| \sum_{qr \in E_{(9,9)}(F)} \frac{2\sqrt{S_q S_r}}{S_q + S_r}.$$

From table 4 and figure 4,

$$= 6\frac{2\sqrt{48}}{14} + 3\frac{2\sqrt{64}}{16} + 6\frac{2\sqrt{72}}{17} + \frac{3^{n++1} - 33}{2} \times \frac{2\sqrt{81}}{18}$$

$$GA_5(H_n) = \frac{12^{3/2}}{7} + 3 + \frac{72\sqrt{2}}{17} + \frac{3^{n+1} - 33}{2}.$$

Here the proof of the theorem 12 is completed.

## 4. Conclusion

The problem of finding the general formula for edge version of $ABC$ index, $ABC_4$ index, Randic connectivity index, sum connectivity index, $GA$ index and $GA_5$ index of Double-wheel graph $DW_n$ and Hanoi graph $H_n$ is solved here analytically.

## References

[1] Chen, J., & Li, S. (2011). On the sum-connectivity index of unicyclic graphs with k pendent vertices. *Mathematical Communications*, *16*(2), 359-368.

[2] Das, K. C., & Trinajstić, N. (2010). Comparison between first geometric–arithmetic index and atom-bond connectivity index. *Chemical physics letters*, *497*(1), 149-151.

[3] Estrada, E., Torres, L., Rodríguez, L., & Gutman, I. (1998). An atom-bond connectivity index: modelling the enthalpy of formation of alkanes. *Indian journal of chemistry. Sect. A: Inorganic, physical, theoretical & analytical*, *37*(10), 849-855.

[4] Farahani, M. R. (2012). Some connectivity indices and Zagreb index of polyhex nanotubes. *Acta Chim. Slov*, *59*, 779-783.

[5] Farahani, M. R. (2013). Computing fourth atom-bond connectivity index of V-Phenylenic Nanotubes and Nanotori. *Acta Chimica Slovenica*, *60*(2), 429-432.

[6] Farahani, M. R. (2014). Computing GA_ {5} index of armchair polyhex nanotube. *Le Matematiche*, *69*(2), 69-76.

[7] Ghorbani, M., & Hosseinzadeh, M. A. (2010). Computing ABC4 index of nanostar dendrimers. *Optoelectron. Adv. Mater. Rapid Commun*, *4*, 1419-1422.

[8] Gan, L., Hou, H., & Liu, B. (2011). Some results on atom-bond connectivity index of graphs. *MATCH Commun. Math. Comput. Chem*, *66*(2), 669-680.

[9] Graovac, A., Ghorbani, M., & Hosseinzadeh, M. A. (2011). Computing fifth geometric-arithmetic index for nanostar dendrimers. *J. Math. Nanosci*, *1*(1), 32-42.

[10] Kanna, M. R., Kumar, R. P., & Jagadeesh, R. (2016). Computation of Topological Indices of Dutch Windmill Graph. *Open Journal of Discrete Mathematics*, *6*(02), 74.

[11] Randic, M. (1975). Characterization of molecular branching. *Journal of the American Chemical Society*, *97*(23), 6609-6615.

[12] Sridhara, G., Kanna, M. R., & Indumathi, R. S. (2015). Computation of topological indices of graphene. *Journal of Nanomaterials*, *16*(1), 292.

[13] Shigehalli, V. S., & Kanabur, R. (2016). Computation of New Degree-Based Topological Indices of Graphene. *Journal of Mathematics*, *2016*.

[14] Vukičević, D., & Furtula, B. (2009). Topological index based on the ratios of geometrical and arithmetical means of end-vertex degrees of edges. *Journal of mathematical chemistry*, *46*(4), 1369-1376.

[15] Zhou, B., & Trinajstić, N. (2009). On a novel connectivity index. *Journal of mathematical chemistry*, *46*(4), 1252-1270.